\documentclass[aps,prl,amsmath,amssymb,floatfix,twocolumn,superscriptaddress]{revtex4-1}
\usepackage{graphicx}
\usepackage{subfigure}
\usepackage{color}
\usepackage{hyperref}
\usepackage{gensymb}

\begin{document}

\title{Optical activity as a test for dynamic chiral magnetic effect of Weyl semimetals}
\author{Pallab Goswami}
\affiliation{National High Magnetic Field Laboratory, Florida State University, Tallahassee, Florida 32310, USA}
\affiliation{Condensed Matter Theory Center and Joint Quantum Institute, University of Maryland, College Park, Maryland 20742-4111, USA}

\author{Girish Sharma}
\affiliation{Department of Physics and Astronomy, Clemson University, Clemson, SC
29634, USA}
\author{Sumanta Tewari}
\affiliation{Department of Physics and Astronomy, Clemson University, Clemson, SC
29634, USA}

\begin{abstract}
Recent angle resolved photoemission spectroscopy measurements have identified an inversion symmetry breaking Weyl semimetal phase in TaAs and NbAs. In an inversion symmetry breaking Weyl semimetal the left and the right handed Weyl points can occur at different energies and the energy mismatch between the Weyl points of opposite chirality is known as the chiral chemical potential. In the presence of the chiral chemical potential, the nontrivial Berry curvature of the Weyl fermions gives rise  to the \emph{dynamic} chiral magnetic effect. This describes how a time dependent magnetic field leads to an electrical current along the applied field direction, which is also proportional to the field strength. We derive a general formula for the dynamic chiral magnetic conductivity of the inversion symmetry breaking Weyl semimetal. We show that the measurement of the natural optical activity or rotary power provides a direct confirmation of the existence of the dynamic chiral magnetic effect in inversion symmetry breaking Weyl semimetals.
\end{abstract}

\maketitle

{\em Introduction :} Recently there has been a surge of interest in three dimensional Weyl semimetals (WSM), which are gapless systems with nontrivial momentum space topology. In these systems two non-degenerate bands linearly cross at isolated points in the momentum space, which act as the unit strength monopole and anti-monopole of the Berry curvature $\Omega_{n,\mathbf{k}}$ of the Bloch bands indexed by $n$ \cite{Volovik, Wan}. The low energy quasiparticles around the monopole and the antimonopole of the Berry curvature are respectively known as the right and the left handed Weyl fermions and the topological invariant of the Weyl fermions is entirely specified by the strength of the monopoles. The existence of the nontrivial Berry curvature requires that the spatial inversion ($\mathcal{P}$) and/or the time reversal ($\mathcal{T}$) symmetries are broken. In the past few years there have been many theoretical proposals for realizing the Weyl quasiparticles in semimetal \cite{Burkov1,Burkov2,Zyuzin1,Cho,Liu,Das} and superconducting \cite{Gong,Sau,Meng,GoswamiBalicas,GoswamiNevidomskyy} systems. On the experimental side researchers have only recently identified the existence of an inversion symmetry breaking WSM phase~\cite{Bernevig,Zahid1} in TaAs~\cite{Zahid2,Lv} and NbAs~\cite{Zahid3} through angle resolved photoemission experiments. Because of $\mathcal{T}$ symmetry, an inversion asymmetric WSM can only support an even number of right and left handed pairs of Weyl fermions, in contrast to WSMs with broken $\mathcal{T}$. Thus, while a non-zero anomalous Hall effect serves as the characteristic topological  magneto-electric effect (mediated by non-trivial Berry curvature) in WSMs with broken $\mathcal{T}$ symmetry, in the inversion symmetry breaking WSMs the anomalous Hall effect identically vanishes. In the inversion asymmetric WSMs, on the other hand, the energy difference between the right and the left handed Weyl fermions, also known as the chiral chemical potential, can be naturally non-zero. In this paper we propose an experimental test for the dynamic chiral magnetic effect (DCME), which is an intriguing topological magneto-electric effect of inversion symmetry breaking WSM, through the measurement of natural optical activity or rotary power of the transmitted light.

The nontrivial Berry curvature of WSM can lead to many anomalous transport properties. For simplicity, let us consider a WSM with one pair Weyl points of opposite chirality, which are separated in the momentum space by $\delta\mathbf{K}$ and in energy space by $\Delta$ (also known as the chiral chemical potential). Such a system breaks both $\mathcal{T}$ and $\mathcal{P}$ symmetries. Field theoretic calculations suggest that such a system possesses a dynamic magneto-electric coupling $\theta(\mathbf{x},t) \mathbf{E} \cdot \mathbf{B}$, where $\theta(\mathbf{x},t)=(\delta\mathbf{K}\cdot \mathbf{x}-\Delta  \; t)$~\cite{Zyuzin3,Grushin,GoswamiTewari2}. Since $\delta \mathbf{K}$ and $\Delta$ are respectively odd under time reversal and inversion operations ($\delta \mathbf{K} \rightarrow -\delta \mathbf{K}$ and $\Delta \to -\Delta$), the magnetoelectric coefficient $\theta(\mathbf{x}, t)$ is $\mathcal{P}$ and $\mathcal{T}$ odd. This magnetoelectric coupling leads to the anomalous charge current
\begin{equation}
\mathbf{j}=\frac{e^2}{h} \delta \mathbf{K} \times \mathbf{E} + \frac{e^2}{h^2} \Delta \; \mathbf{B}.
\end{equation}

In the absence of $\mathcal{T}$ symmetry, the spatial gradient of $\theta(\mathbf{x},t)$ governs the anomalous charge Hall current. In contrast, the time derivative of $\theta(\mathbf{x},t)$ induces a current along the direction of the applied magnetic field for the inversion asymmetric Weyl semimetal, which is known as the {\em chiral magnetic current} (CME)\cite{Kharzeev,Landsteiner}. Recently, there has been some controversy regarding the existence of the equilibrium CME in condensed matter systems \cite{Franz,Basar,Landsteiner2,Burkov4,GoswamiTewari1}. In contrast, it is accepted that a DCME can exist in response to a time dependent magnetic field. In particular, the existence of the DCME has been argued in the quark-gluon plasma \cite{Kharzeev,Kharzeev1}, and rotating $^3He-A$ \cite{Volovik}. Therefore it is natural to ask about the possible experimental ramifications of the DCME in solid-state realizations of inversion symmetry breaking WSM. This question assumes even more importance in light of the recent experimental observations of the inversion symmetry breaking WSM, for which the anomalous Hall conductivity vanishes due to the presence of $\mathcal{T}$ symmetry, and DCME (that requires the existence of the inversion symmetry breaking chiral chemical potential) is the
characteristic signature of the topological magneto-electric effect.

In this paper we show that the DCME current is intimately related to the natural optical activity of an inversion symmetry breaking metal, which is also known as the optical gyrotropy \cite{Landau,Mineev1, Moore, Raghu1, Mineev2, Chakravarty}. The optical gyrotropy arises due to the existence of the gyrotropic current in response to the time derivative of the magnetic field or the curl of the electric field, which are related by the Maxwell's equation, $\partial_t \mathbf{B}=-\nabla \times \mathbf{E}$. We employ the following definitions of the chiral magnetic and the gyrotropic currents
\begin{eqnarray}
\mathbf{j}_{ch}(\mathbf{q},\omega)=\sigma_{ch}(\mathbf{q},\omega) \; \mathbf{B}(\mathbf{q},\omega),\label{eq3}\\
\mathbf{j}_{g}(\mathbf{q},\omega)=\sigma_{g}(\mathbf{q},\omega) \; i\omega\; \mathbf{B}(\mathbf{q},\omega)\label{eq4}
\end{eqnarray}
where $\sigma_{ch}(\mathbf{q},\omega)$ and $\sigma_{g}(\mathbf{q},\omega)$ are respectively the complex chiral magnetic and the complex gyrotropic conductivities. The real and the imaginary parts of the $\sigma_{ch}(\mathbf{q},\omega)$ respectively lead to the currents, which are in and out of phase with the magnetic field. By comparing two definitions we immediately find the remarkable relation
\begin{equation}
\sigma_{ch}(\mathbf{q},\omega)=i \omega \; \sigma_{g}(\mathbf{q},\omega).
\end{equation}
In the presence of a nonzero gyrotropic conductivity, the refractive indices for the left and the right circularly polarized light become different, which in turn causes a rotation of the plane of polarization for the transmitted light, even in the absence of a uniform external magnetic field. The amount rotation of the plane of polarization per unit length is known as the rotary power, and the imaginary part of the gyrotropic conductivity governs the size of the rotary power~\cite{Landau}. Therefore, the central result of this paper is that the natural optical activity can be used as a test for the existence of the real part of the dynamic chiral magnetic conductivity of inversion symmetry breaking WSMs such as TaAs and NbAs. Below we derive a formula for the dynamic chiral magnetic conductivity and optical activity (rotary power) of an inversion asymmetric WSM. \\


{\em Tight-binding model of inversion asymmetric WSM:} Experimental systems such as TaAs and NbAs are body centered tetragonal systems which possess twelve pairs of right and left handed Weyl fermions (altogether twenty four Weyl points), and currently there is no simple tight binding model for describing these materials. For this reason we adopt a simple two band model defined on a cubic lattice, which can easily capture all the physically interesting topological aspects of an inversion symmetry breaking WSM. Our Hamiltonian is defined as
\begin{equation}
H=\sum_{\mathbf{k}} \psi^\dagger_{\mathbf{k}}\left[N_{0,\mathbf{k}} \; \sigma_0 + \mathbf{N}_{\mathbf{k}} \cdot \boldsymbol \sigma \right]\psi_{\mathbf{k}}, \label{tightbinding}
\end{equation}
where the spin independent hopping term $N_{0,\mathbf{k}}$ and the spin-orbit coupling terms $\mathbf{N}_{\mathbf{k}}$ are respectively even and odd under spatial inversion and the Pauli matrices $\boldsymbol \sigma$ operate on the physical spin indices. For simplicity we only choose nearest neighbor hopping terms and obtain $\mathbf{N}_{\mathbf{k}}=t_{SO}[\sin (k_1a), \sin (k_2a), \sin (k_3a)]$, and $N_{0,k}=-2t_1[\cos (k_1a) +\cos (k_2a) +\cos (k_3a)]$, with $t_1 \ll t_{SO}$, where $a$ is the lattice spacing. The quasiparticle spectra are given by
\begin{equation}
E_{n, \mathbf{k}}= N_{0,\mathbf{k}} + (-1)^n |\mathbf{N}_{\mathbf{k}}|,
\end{equation}where $n=1,2$. The Berry curvatures for the Bloch bands are given by
\begin{equation}
\Omega_{n,\mathbf{k},a}= (-1)^{n+1} \Omega_a= \frac{\epsilon_{abc}}{4N^3} \; \mathbf{N} \cdot (\partial_b \mathbf{N} \times \partial_c \mathbf{N}).
\end{equation} The Weyl excitations occur around the eight high symmetry points $\Gamma$: (0,0,0); M: ($\pi/a$, $\pi/a$, 0), ($\pi/a$, 0, $\pi/a$), (0, $\pi/a$, $\pi/a$); X: ($\pi/a$, 0,0), (0, $\pi/a$, 0), (0, 0, $\pi$); and R: ($\pi/a$, $\pi/a$, $\pi/a$) of the cubic Brillouin zone. In addition, $\Gamma$ and $M$ points act as the four right handed Weyl points while $R$ and $X$ points are the four left handed Weyl points. In the vicinity of these points, quasiparticles possess linear dispersion and the Berry curvature  acquires the characteristic form of a monopole (antimonopole)
$$\Omega_{a}= (-1)^{(K_1+K_2+K_3)} \; \mathrm{sgn}(t_{SO}) \;  \frac{k_a}{2k^3},$$ where $K_j$s are the components of the wave-vectors of the Weyl points. The spin independent hopping term shifts the Weyl points in the energy space, and gives rise to the chiral chemical potential. The linearized Hamiltonian for the low energy quasiparticles can be compactly written as
\begin{equation}
H \approx \sum_j \int \frac{d^3k}{(2\pi)^3} \psi^\dagger_{j,\mathbf{k}} [\Delta_j + \sum_{b=1}^{3} v_{b,j} (k_a-K_{b,j}) \sigma_b]\psi_{j,\mathbf{k}},
\label{eq:cont}
\end{equation} where the momentum deviation from the Weyl points are restricted by an ultraviolet cutoff $\Lambda \sim \pi/a$, and the velocity components satisfy $|v_{b,j}|=|t_{SO}| a/\hbar$. Specifically the crossing of two bands at $\Gamma$, $M$, $X$ and $R$ respectively occur at energies $\Delta_j=-6t_1, \; 2t_1, \; -2t_1, \;6t_1$. If overall chemical potential is set to zero this system will behave as a compensated semimetal. If $t_1>0$, we have electron pockets centered at $\Gamma$ and $X$ points and hole pockets around $M$ and $R$ points. By choosing further neighbor spin independent hopping the reference energies and the volumes of the Fermi pockets can be modified. Therefore our simple tight binding model clearly explains how the energy mismatch between the Weyl points can naturally arise in a material.
\\

{\em Dynamic chiral magnetic conductivity :} In the presence of the electromagnetic field we need to replace $\mathbf{k}$ by $\mathbf{k} -e \mathbf{A}$ and also consider a dynamic Zeeman coupling $g \mu_B \boldsymbol \sigma \cdot \nabla \times \mathbf{A}$, which leads to the following current operators
\begin{equation}
j_a(\mathbf{k})=-e [\partial_a N_{0,\mathbf{k}}+ \partial_a \mathbf{N}_{\mathbf{k}} \cdot \boldsymbol \sigma + i\gamma \epsilon_{abc} \sigma_b k_c],
\end{equation} and $\gamma=g \mu_B/e$. For simplicity, we will focus on the high frequency regime ($\omega \tau >>1$), where the scattering effects due to impurities can be ignored. In this collisionless regime, the chiral magnetic conductivity can be calculated using the following Kubo formula~\cite{Kharzeev,Landsteiner}
\begin{equation}\label{eq10}
\sigma_{ch}(\mathbf{q},\omega)=\frac{\epsilon_{abc}}{2iq_c} \; \Pi_{ab}(i\Omega_m \to \omega + i \delta, \mathbf{q}),
\end{equation} where
\begin{eqnarray}
\Pi_{ab}(i\Omega_m,\mathbf{q})&=&T \sum_{n} \int \frac{d^3k}{(2\pi)^3} Tr[j_{a,\mathbf{k}}G(i\omega_n,\mathbf{k}-\mathbf{q}/2) \nonumber \\ && \times j_{b,\mathbf{k}}G(i\omega_n+i\Omega_m,\mathbf{k}-\mathbf{q}/2)]
\end{eqnarray} is the current-current correlation function and
\begin{equation}
G(i\omega_n,\mathbf{k})=\frac{i\omega_n+\mu-N_{0,\mathbf{k}}+\mathbf{N}_{\mathbf{k}}\cdot \boldsymbol \sigma}{(i\omega_n+\mu-N_{0,\mathbf{k}})^2-|N_{\mathbf{k}}|^2}
\end{equation}
is the propagator for the fermions. In the above equations, $\omega_n=(2n+1)\pi T$ and $\Omega_m=2\pi m T$ are respectively the fermionic and the bosonic Matsubara frequencies, and $n$, $m$ are integers.

For long wavelength electromagnetic waves, $\omega \gg v_F |\mathbf{q}|$ and we can ignore the $q$ dependence of the chiral magnetic conductivity. This approximation also provides the answer for the chiral magnetic conductivity in the presence of a time dependent, but spatially homogeneous magnetic field. Therefore, we restrict ourselves to finding the $q$-linear part of $\Pi_{ab}$. In the following we choose $\Pi_{12}$ as a function of $\mathbf{q}=(0,0,q)$. After performing the trace and the Matsubara sum we obtain the following two contributions to the dynamic chiral magnetic conductivity
\begin{widetext}
\begin{eqnarray}
\sigma_{ch,1}(i\omega_m)&=& 2e^2\gamma^2 \sum_{n=1}^{2} \; \int_{\mathbf{k}} \left[ \frac{n^{\prime}_F(E_n) \; \mathbf{k} \cdot \mathbf{N}_{\mathbf{k}} \; k_3 \partial_3E_n}{T[(i\omega_m)^2-4|\mathbf{N}_{\mathbf{k}}|^2]}-\frac{(-1)^n n_F(E_n)\; \mathbf{k} \cdot \mathbf{N}_{\mathbf{k}} \; k_3 \partial_3N_{0,\mathbf{k}}}{|\mathbf{N}_{\mathbf{k}}|[(i\omega_m)^2-4|\mathbf{N}_{\mathbf{k}}|^2]}\left \{1-\frac{2(i\omega_m)^2}{(i\omega_m)^2-4|\mathbf{N}_{\mathbf{k}}|^2} \right \}\right], \label{ch1} \nonumber \\
&& \\
\sigma_{ch,2}(i\omega_m)&=&4e^2 \sum_{n=1}^{2} \; \int_{\mathbf{k}} |\mathbf{N}_{\mathbf{k}}|^3 \bigg[\frac{n^{\prime}_F(E_n)  \; \Omega_{3,\mathbf{k}} \partial_3E_n }{T[(i\omega_m)^2-4|\mathbf{N}_{\mathbf{k}}|^2]}-\frac{(-1)^n n_F(E_n)\; \nabla N_{0,\mathbf{k}} \cdot \mathbf{\Omega}_{\mathbf{k}}}{|\mathbf{N}_{\mathbf{k}}|[(i\omega_m)^2-4|\mathbf{N}_{\mathbf{k}}|^2]}+\frac{(-1)^n  n_F(E_n)\; (i\omega_m)^2 }{|\mathbf{N}_{\mathbf{k}}|[(i\omega_m)^2-4|\mathbf{N}_{\mathbf{k}}|^2]^2} \nonumber \\ && \times 2 \Omega_{3,\mathbf{k}} \partial_3 N_{0,\mathbf{k}}  \bigg],\label{ch2}
\end{eqnarray}
\end{widetext}which respectively arise from the dynamic Zeeman coupling and the underlying Berry curvature. These two equations constitute our main result, and can also be applied to a generic inversion symmetry breaking metal. After an analytic continuation to the real frequencies, we obtain the complex chiral magnetic conductivity.

Now we apply these formula for evaluating the chiral magnetic conductivity of a WSM with a finite chiral chemical potential. For simplicity now we use the continuum Hamiltonian, as described by Eq.~(\ref{eq:cont}). 
The entire expression now depends on the derivative of the Fermi function, which at zero temperature reduces to a delta function, and the dynamic chiral magnetic conductivity for this model becomes
\begin{eqnarray}\label{eq16}
\sigma_{ch}(\omega)&=&\frac{e^2 \mathrm{sgn}(t_{SO})}{3\pi^2\hbar^2} \sum_j (-1)^{(K_{1,j}+K_{2,j}+K_{3,j})} \nonumber \\
&& \times \left(1+\frac{\gamma^2\mu^2_j}{\hbar^2|v|^4}\right)\frac{\mu^3_j}{( \hbar \omega+i\delta )^2-4\mu^2_j} ,
\end{eqnarray}where $\mu_{j}=\mu + \Delta_j$, and $\mu$ is the conventional chemical potential, $|v|=|t_{SO}|a/\hbar$. 
Notice that the imaginary part of the $\sigma_{ch}$ only appears when the frequency is tuned exactly at the effective chemical potentials $\mu_{j}$ of the Weyl quasiparticles. Even though we obtain a finite chiral magnetic conductivity by setting $\omega=0$ in Eq.~(\ref{eq16}), this answer does not constitute an equilibrium CME. For the equilibrium CME, we have to first set $i\omega_m$ to be zero in Eq.~(\ref{eq10})~\cite{GoswamiTewari1}. On a general ground we anticipate the low frequency dynamic chiral magnetic conductivity obtained from Eq.~(\ref{ch1}) and Eq.~(\ref{ch2}) to be considerably modified by various relaxation processes. In contrast, there is an universality in the high frequency regime. At frequencies much higher than the chemical potentials of the Weyl fermions (which is quite small in TaAs and NbAs) and the scattering rate $1/\tau$, the chiral magnetic conductivity is entirely real and decreases as $1/\omega^2$. The corresponding asymptotic form is given by
\begin{eqnarray}
\sigma_{ch}(\omega) & \sim & \frac{e^2 \mathrm{sgn}(t_{SO})}{3\pi^2\hbar^2}\sum_j (-1)^{(K_{1,j}+K_{2,j}+K_{3,j})} \frac{\mu^3_j}{(\hbar \omega)^2} \nonumber \\ & & \times \left(1+\frac{\gamma^2\mu^2_j}{\hbar^2|v|^4}\right).
\end{eqnarray} This equation also implies that the high frequency gyrotropic conductivity is purely imaginary and decreases as $1/\omega^{3}$.

Before determining the overall magnitude of the chiral magnetic conductivity, we first compare the strengths of the contributions from the dynamic Zeeman coupling and the Berry curvature. This is quantified by the dimensionless
ratio $\gamma^2 \mu^2_j/(\hbar^2 v^4)$. Recent magnetotransport experiments on TaAs~\cite{Jia} suggest the following estimates $\mu \sim 11.48 meV$, $v \sim 1.16 \times 10^5 m/s$, and $k_F=\mu/(\hbar v) \sim 1.5 \times 10^8 m^{-1}$. For simplicity we choose $g=2$ and $\gamma=\hbar/m_e$, where $m_e$ is electron's rest mass. Therefore, the dimensionless ratio becomes
\begin{eqnarray}
\frac{\gamma^2 \mu^2}{\hbar^2 v^4} = \frac{\mu^2}{m^2_e v^4}=\frac{\hbar ^2 k^2_F}{m^2_e v^2} \sim 1.9 \times 10^{-2}.
\end{eqnarray} This estimate suggests that the contribution from the Berry curvature dominates over the one from the dynamic Zeeman coupling if $g \sim 2$. However, in a semiconducting system the $g$ factor can be substantially different from $2$. In particular, along certain directions it can be an order of magnitude larger than $2$. Such an enhanced $g$ factor can make both contributions comparable.

Now we turn to the estimation of the chiral magnetic conductivity due to the Berry curvature. When the frequency of the electromagnetic field is bigger than the plasma frequency of the Weyl fermions $\omega_p \sim |\mu|/\hbar$~\cite{Hwang,LvZhang} and the scattering rate $\tau^{-1}$ due to impurities, the chiral magnetic conductivity can be approximated as
\begin{equation}
\sigma_{ch}(\omega) \sim  \frac{128 e^2 t^3_1}{\pi^2 \hbar^4 \omega^2}.\label{chestimation}
\end{equation} For TaAs, the $\hbar \omega_p \sim 11.48 meV$ and $\hbar/\tau \sim 5 meV$, and we can apply Eq.~(\ref{chestimation}) for the estimation of $\sigma_{ch}$ if $\hbar \omega >> 11.5 meV$. In particular, the infrared frequencies $50 meV <\hbar \omega< 1.7 eV$ will be suitable for experimental detection of chiral magnetic conductivity of different WSMs. In the presence of the chiral magnetic conductivity, the right and the left circularly polarized electromagnetic waves propagate with different velocities, and also possess different refractive indices~\cite{Landau,Jackiw}. This gives rise to the natural optical activity which is described in terms of the rotary power (rotation of the plane of polarization of the transmitted light per unit length)~\cite{Landau, Raghu1, Hosur}
\begin{equation}
\frac{d\theta}{dl}=\frac{h}{2e^2c}\Re[\sigma_{ch}(\omega)] .
\end{equation}
where $c$ is the speed of light in the vacuum. Therefore, the rotary power, which is a very well known and widely measured quantity can provide a direct measurement of the real part of the chiral magnetic conductivity. By choosing $t_1 \sim 5 meV$, we find that $\frac{d\theta}{dl}$ decreases from $665 \degree /mm $ (for $\hbar \omega =50 meV$) to $0.6 \degree /mm$ (for $\hbar \omega =1.7 eV$). For $\hbar \omega= 62 meV$ our estimated rotary power $424 \degree /mm$ is about eight times larger than the experimentally observed rotary power $\sim 50 \degree /mm$ of tellurium~\cite{NomuraTe,BrownTe}. As the rotary power of tellurium exceeds that of quartz for infrared frequencies~\cite{NomuraTe}, we can safely conclude that inversion asymmetric WSMs can generically display a large rotary power. It is important to note that the natural optical activity in the presence of time reversal symmetry does not lead to the polar Kerr effect, which describes the rotation of the plane of polarization for the reflected light~\cite{Raghu2, Hosur}. Recently in Ref.~\onlinecite{Hosur}, the authors have considered the possibility of enhancing the optical activity of the WSM by applying external electric and magnetic fields in parallel, which can serve as a potential test of the chiral anomaly. Their results are somewhat different from ours, as they have not considered an equilibrium chiral chemical potential, and the chiral chemical potential rather emerges as a consequence of the chiral anomaly (and it is proportional to applied $\mathbf{E}\cdot\mathbf{B}$). In addition, the authors have not considered the role of the Zeeman coupling either.

In conclusion, we have established a direct relation between the DCME and the optical gyrotropy for inversion symmetry breaking WSMs (see Eq.~(\ref{eq4})). Based on this observation we have suggested the natural optical activity or the rotary power for the transmitted light as a direct evidence for the existence of the DCME for inversion asymmetric WSMs such as TaAs and NbAs. We have introduced a simple tight binding model on a cubic lattice, which produces eight Weyl points with different reference energies or chiral chemical potentials. Our general formula for the high frequency chiral magnetic conductivity (see Eq.~(\ref{ch1}) and Eq.~(\ref{ch2})), can also be applied to generic inversion symmetry breaking metals such as Li$_2$Pt$_3$B and Li$_2$Pd$_3$B~\cite{Pickett,Takimoto}.\\

{\em Acknowledgment:} This work is supported by the NSF Cooperative Agreement No.DMR-
0654118, the State of Florida, and the U. S. Department
of Energy (P.G.) and NSF (PHY-1104527) and AFOSR (FA9550-13-1-0045) (G. S. and S.T.). We thank S. Chakravarty and S. Raghu for pointing out the reason why the natural optical activity in the absence of time reversal symmetry breaking does not lead to the polar Kerr effect, and drawing our attention to Ref.~\onlinecite{Halperin}.

\end{document}